\documentclass[12pt]{article}
\usepackage{graphicx}

\title{\bf }
\author{\it B.F. Kostenko, M.Z. Yuriev}
\begin{document}

\begin{center}
\Large{\bf Possibility of a modification of life time of
radioactive elements by magnetic monopoles}\\
\smallskip
\large{\it B.F. Kostenko}\\
\smallskip
\small{Joint Institute for Nuclear Research, Dubna}\\ \large{\it
M.Z. Yuriev}\\
\smallskip
\small{EuroFinanceGroup, Moscow}\\

\smallskip

\end{center}




\section{Introduction}

The existence of magnetic charges has appeal from the theoretical
point of view: it explains the quantization of the electric charge
and symmetrize Maxwell's equation. Therefore, from 1931, when the
famous Dirac paper \cite{Dirac} was published,  searches for
magnetic monopoles were carried out at every new generation of
accelerators, but all those attempts were futile. Now it is
generally accepted  that magnetic monopoles, if they exist, should
be very heavy, with mass $\geq$ 500--1000 GeV. Nevertheless, an
interesting possibility of existence of relatively light magnetic
charges follows from G. Lochak's magnetic monopole concept
\cite{Lochak1}--\cite{Lochak3}. Indeed, all accelerator magnetic
monopole searches are based on the existence of the vertex $\gamma
\rightarrow M + \overline{M}$, suggested by analogy with $\gamma
\rightarrow e^- + e^+ .$ The most straightforward way to prevent
the monopole--antimonopole creation in the accelerator experiments
is to permit the following violation of C invariance in the
electromagnetic interactions\footnote{We believe that Prof. G.
Lochak will agree with our interpretation of his theory. A
possibility of violation of C invariance in the electromagnetic
interactions was supposed earlier, although in a different
physical context, by T.D. Lee et al. \cite{Ryder}.}: it is
possible to assume that antimonopole, $\overline{M}$,
corresponding to the solution of the Dirac equation with negative
energy, does not take part in the electromagnetic interactions (in
contrast to $ e^+$). This means that the vertex $\gamma
\rightarrow M + \overline{M}$ does not exist, in spite of the
existence of the vertex $M \rightarrow M + \gamma$.

There is a close analogy of such a violation of C invariance with
P violation in the weak interaction. In  the latter case $\nu_R$
and $\overline{\nu}_L$ do not take part in the $V-A$ interaction
(are "sterile" particles). The only difference refers to the
question of the existence of the particles. Since any interactions
of  $\nu_R$ and $\overline{\nu}_L$ are unknown, the very existence
of these particles is still doubtful, whereas the existence of
negative energy particles follows from the requirement of the
possibility of spatial localization of the positive energy
solution \cite{Bjorken}. The monopoles may be even massless (the
linear variant of the Lochak theory
\cite{Lochak1}--\cite{Lochak3}). In this case positive and
negative energy monopoles (interacting and "sterile" ones)  are
present in the wave packet in the equal ratio, in the complete
analogy with the massless neutrino field. Decay of the vacuum to
monopole-antimonopole pairs, possible in principle only through
the chain $$ | 0 \rangle \rightarrow \gamma + M + \overline{M}
\rightarrow M + \overline{M} $$ is, in fact, forbidden. Indeed,
the vertex $ \gamma + M + \overline{M}$ with a virtual $\gamma$,
which is absorbed by $M$ in a subsequent moment of time, is
blocked due to the sterility of $\overline{M}$.

In papers \cite{Lochak1}--\cite{Lochak3}, some heuristic
arguments, based on a macroscopic gedanken experiment, are given
in favour of generalization of C, P, T operations on the case of
observable particles with different helicities, which remain in
the theory after "deleting" the negative energy states from the
Dirac spinor.

Since Lochak's monopoles are unregistered in the accelerator
experiments, two interconnected problems arise: to formulate a
theory describing monopole production (it should include a new
force beyond the Standard model of electroweak interaction), and
to point the way to monopole observation. G. Lochak et al. assumed
these monopoles to be produced by strong magnetic pulses inside
atomic nuclei able to the weak decays (see \cite{Lochak3} for the
references). In the present paper we consider a possibility,
closely related to these and some other experiments, of a
modification of life time of radioactive elements by magnetic
monopoles. The first part of the article is devoted to purely
electromagnetic impact of monopoles, caused by the vertex $M
\rightarrow M + \gamma$. The second part, more speculative one, is
based on experimental evidences in favour of the existence of some
axial vector currents, responsible for a new force, which can
stimulate, or suppress, decays of radioactive elements.

\section{Electromagnetic interactions}

It is common knowledge that a possibility of $\beta$-decays into
bound states begins to play a crucial role if the energy released
in this process is comparable with binding energies of electrons
in the atom. Experiments demonstrate a great difference, up to
nine orders, between the decay rates of neutral atoms and their
totally bare ions \cite{GSI1}, \cite{GSI2}. These results are
quite clear from the general formula for probability of quantum
transition, $$ \lambda = 2\pi \sum_f |<f|H_{int}|i>|^2 \delta (E_f
- E_i),$$ since in the above-mentioned experiments the final phase
space is substantially extended after ionization even only one
energy level, $|f>$, which can be occupied by an outgoing
electron.

The decay rate of radioactive atoms placed in an external magnetic
field should  vary too if the field is strong enough to modify the
number of allowed final states. Thereupon it should be noted that
a weak magnetic field responsible for the Zeeman, or Paschen-Back
splitting does not change the number of states which could be
simultaneously occupied by electrons in atom (this splitting can
be observed only in atomic spectra). Otherwise already the Earth
magnetic field should lead to a disaster, bringing down all
electrons' orbits. Nevertheless, the radioactive atom create one
more vacancy which can be occupied by an electron produced in
$\beta$ decay due to the increase of the nuclear charge, $Z
\rightarrow Z+1,$ in these processes. The probability of these
transitions is proportional to the density of unoccupied electron
levels in the vicinity of the nucleus. In the absence of a
magnetic field, the density of excited electron orbit at the
position of the nucleus decreases very fast with increasing the
principal quantum number $n_p$, in proportion to $1/n_p^3$, and
the probability of $\beta$-decay with a small energy release is
really tiny. The present state of affairs changes drastically if
the $\beta$-radioactive atom is placed into a strong magnetic
field.

Loudon was the first who considered the behaviour of atomic
electron in a very strong magnetic field \cite{Loudon}. In this
case the energy of magnetic interaction begins to dominate over
the Coulomb one, atom acquires an elongated shape along the
magnetic field with the transverse spread much smaller than the
longitudinal extent, and one has a quasi-one-dimensional atom with
Coulomb interaction\footnote{It is interesting to note that the
{\it genuine} one-dimensional Coulomb problem has a solution
\cite{one-dim} different from that found by Loudon, but that
solution is not so important from the physical point of view. In
fact, Loudon investigated electron with a small effective mass in
the matter with a big dielectric constant. This increases the
magnetic interaction energy and decreases the Coulomb one so that
a laboratory field of $2.4\times 10^4 $ G corresponds to an
effective magnetic field of $3.6\times 10^{10} $ G.} $$ V(z) = -
e^2/ |z|.$$ In the high magnetic field regime, one is dealing with
the motion of almost free electrons in a magnetic field. The
corresponding physical conditions can be expressed in different
equivalent forms:
\begin{equation}\label{1}
 \mu_B B = \hbar \omega_L >> Ry,  \qquad r_L = (B_0/B)^{1/2} r_B << r_B,
\end{equation}
where $\mu_B = e\hbar/2m_e c$ and $r_B= \hbar^2/me^2 \approx
0.53\times10^{-8}$ cm are the Bohr magneton and radius, $ \omega_L
= eB/2mc$ and $r_L$ are the Larmor frequency and radius for an
electron moving along a circular orbit in a magnetic field, $Ry =
mc^2 \alpha^2/2 = 13.6$ eV is the Rydberg energy, $B_0 = c m^2 e^3
/\hbar^3 = 2.35 \times 10^9$ G. In this approximation, the wave
function of the electron is simply a product of a Landau wave
function  for the very fast transverse motion, and a function for
the comparatively slow motion parallel to the field.

\begin{figure}
\begin{center}
\includegraphics[width=12cm]{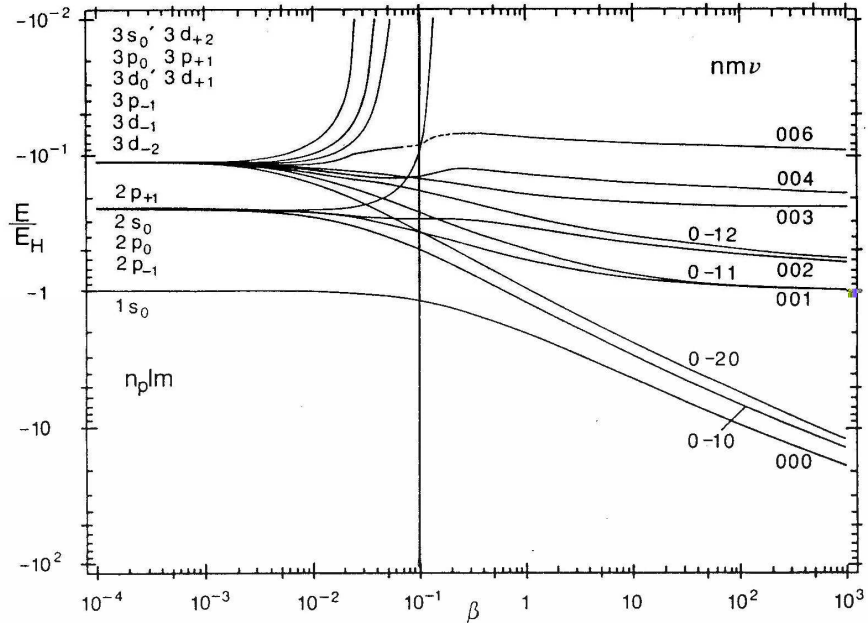}
\end{center}
\caption{\small Energies of the electron in the hydrogen atom as a
function of an external magnetic field field \cite{Ruder} (see
text).}\label{Te}
\end{figure}
In Fig. 1 the energies (in Ry) of low-lying states with principal
quantum number $n_p\leq 3$ for a Coulomb potential are shown as
functions of the magnetic field strength, $\beta = B/2B_0$
\cite{Ruder}. The states at the left, at $\beta<0.1$, are labelled
by atom's field-free quantum numbers, $n_p$, $\it l, \; m $. The
states at $\beta>10$ are enumerated by n (Landau  quantum number),
$\it m$ and $\nu$ (number of nodes of the longitudinal part of the
wave function). One can see that the strong magnetic field not
only increases the number of electron state in atom, but also
decreases the energy of these states due to the electronic orbit
squeezing (in the plane perpendicular to the field) towards the
nucleus. This results in a strong increase of the density of
electron states near the nucleus, which now falls down only as
$1/\nu$ with the increase of the number $\nu$. Estimates show
\cite{Filippov} that the probability of decays into the Landau
levels for an atom containing all its electron, but immersed into
a high magnetic field, should even exceeds the probability of
decay of the totally ionized atom if
\begin{equation}\label{2}
 B/B_0 > 2 Z^2 .
\end{equation}

Of course, the  field induced by the magnetic monopole is
inhomogeneous and one should take this into account. The potential
energy of the Larmor circle in an external magnetic field is:
$$U_L = - \vec \mu \cdot \vec B,$$ where $\vec \mu$ is the
magnetic moment corresponding to the electron circular current in
the plane perpendicular to the external magnetic field, $$ \mu =
\frac{1}{c} I S = \frac{1}{c} \frac{e \omega_L}{2\pi} \pi r_L^2 =
\frac{mv_{\bot}^2}{2B} = \frac{\hbar \omega_L}{2B}, $$ which is
directed against $\vec B,$ $$\vec \mu = - \mu \frac{\vec B}{B}. $$
The magnetic moment is an adiabatic invariant iff the Larmor
radius is sufficiently less than a characteristic scale at which
the magnetic field changes distinctly. For electrons localized
inside atoms, $r\sim r_B$,  one should consider  the adiabatic
condition to be applicable at the scale of about $r_B$. Thus, the
condition takes the form $$r_L << r_B, $$  reproducing exactly the
condition (1).
\begin{figure}
\begin{center}
\includegraphics[width=10cm]{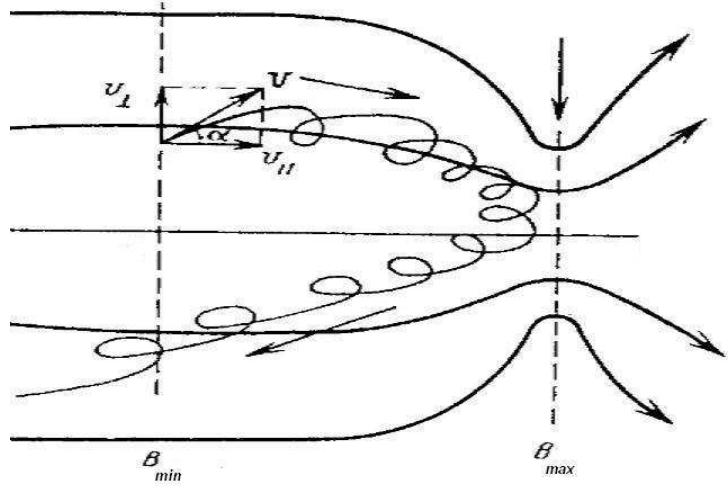}
\end{center}
\caption{\small Plasma mirror-machine \cite{Trubnikov}.
}\label{Mirror}
\end{figure}

The inhomogeneity of the monopole magnetic field gives an
additional force, $$  \vec F = - \vec \nabla U_L = - \mu  \vec
\nabla B,$$ which is well-known in the plasma theory as the
magnetic mirror effect\footnote{Poincar\'e showed \cite{Poincare}
that a trajectory of electric charge moving in a coulombian
magnetic field follows a geodesic line on a cone. Therefore, in
contrast to the plasma mirror-machine shown in Fig.2, a trajectory
of electron bombarding magnetic pole and a trajectory of electron
reflected from it belong to the same cone.} (see Fig.2). This
force is always directed outside from the magnetic charge and
leads to the electron knocking out if
\begin{equation}\label{4}
U_L = \frac{\hbar \omega_L}{2} > {\rm Ionisation \; energy.}
\end{equation}
For external electrons, the ionization energy is of order 1 Ry,
and the condition (3) becomes even weaker than (1), but, of
course, one should always take the strongest of them.

As a preliminary resume, we conclude that the condition (1) is the
weakest for the external magnetic field to influence on the atom
$\beta$-decay. This means that the external magnetic field should
be much stronger than $B_0 = 2.35 \times 10^9$ G. Magnetic
monopole with the minimal charge $g_{\rm min}=68.5 e$ creates, on
the atomic scales $ \sim r_B$, the field of order $10^8$ G. This
implies that only multi-charge monopoles with $g \sim 100 g_{\rm
min}$ could have an essential action upon atom $\beta$-decay
rates.

It is known (see, e.g.,  the right side of Fig.1) that a large
magnetic field increases  the binding energy of electrons in the
atom. As far as the condition of stability of atoms against
$\beta$-decays is given by the condition of the atom  mass minimum
\cite{Mukhin}, \cite{Urutskoev}, the change of  the electron
binding energy  due to the applied magnetic field should also lead
to the change of the $\beta$-decay rate. However, the role of this
effect is much less than those discussed above and it becomes
important if the condition
\begin{equation}\label{5}
\frac{B}{B_0} >> Z^3
\end{equation}
is fulfilled \cite {Filippov}. For $^{187}$Re, this gives $B\sim
10^{15}$ G, or $g\sim 10^{7} g_{\rm min}$.

The increase of the electron binding energy in a strong magnetic
field means that an effective  potential between the atom and
magnetic charge arises. It accelerates atoms toward the monopole
and  could cause  nuclear reactions between them, i.e. could
accomplish a magnetic monopole catalysis. However, the magnetic
mirror force will, of course, depress the process due to the
premature ionization of the atoms.

In the paper \cite{Chernob}, a hypothesis was suggested that the
magnetic monopoles of Georges Lochak type are responsible for the
explosion at the Chernobyl Nuclear Power Plant. According to it,
the magnetic monopoles were formed in the vicinity of turbine
generators and got to the steam pipes. Since the oxygen is
paramagnetic, the magnetic particles formed "bound states" with
oxygen and moved along the steam pipes, together with the steam.
After entering the reactor, the monopoles interacted with
$^{238}$U and, what is prior, with nuclei emitting delay neutrons.

\begin{figure}
\begin{center}
\includegraphics[width=12cm]{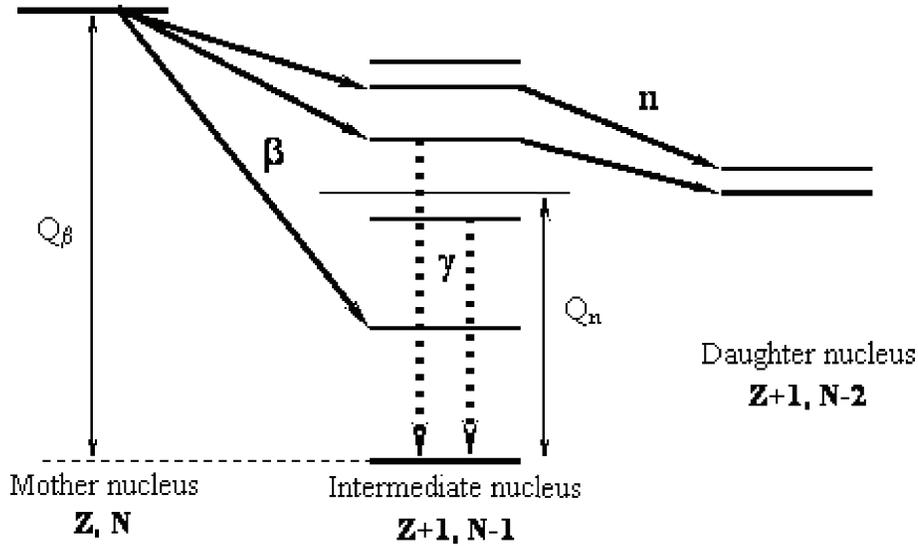}
\end{center}
\caption{\small The scheme of decay of nuclei capable to emit
delayed neutrons. }\label{Delay}
\end{figure}
The scheme of decay of a nucleus emitting a delayed neutron is
shown in Fig.3. Here the mother nucleus, with atomic weight in the
range from A$=$72 to 160, is produced after the fission of
$^{235}$U. The mother nucleus is unstable with respect to the
$\beta$-decay because of an excess of neutrons in it, and the
corresponding $\beta$-transition gives an intermediate nucleus in
an excited nuclear state. If the excitation energy exceeds the
binding energy of neutron, $Q_n$, the intermediate nucleus emits a
neutron. Although this emission takes place practically
instantaneously, a time necessary for the $\beta$-decay is spent
before the delayed neutron is emitted (and this explains the term
"delayed neutron"). The authors have shown that there were about
500 mother nuclei capable of emitting neutrons per each neutron
which was in the reactor at some instant. In the steady-state
regime of reactor operation delayed neutrons amount to only a
small fraction, $\sim 5 \times 10^{-3} $,  of a total number of
neutrons participating in the nuclear fission process at some
instant. But a distortion of the mechanism of decays should
certainly cause a huge increase of the neutron density due to the
huge number of mother nuclei. It was suggested the following
mechanism of the distortion: the magnetic monopoles deformed
electron shells around the mother nuclei. The consequences of the
deformation were much stronger for $\beta$-decays with small
energy release, and, therefore, the number the decays into the
intermediate nuclei capable to emit the delayed neutrons rose
sharply.

According to our previous consideration, such a scenario is
possible only if the monopoles had an unusually large charge $g
\geq 100\; g_{min}$. The following alternative scenario  based on
monopoles with the minimal magnetic charge may be also suggested.
The monopoles, after their creation in the vicinity of turbine
generators, could form bounded states with atoms of steam because
of some kind of attraction between them and the atoms due to an
increase of the electron binding energy of atoms in the strong
magnetic field (as it was discussed above). After penetration in
the reactor, monopoles should be captured by atomic nuclei because
the deepest point of the potential energy is reached there.
Indeed, the monopole with the minimal charge induces the magnetic
field of order $B\sim 10^{17}$ G at distances $r\sim 10^{-13}$ cm.
It gives an energy up to 1 MeV for interaction with the nucleon
magnetic moment $\mu_N = 3.2 \times 10^{-18}$ MeV/G. This means
that intermediate nuclei  may be significantly excited after the
monopole absorption. If the intermediate nuclei  have a high
magnetic momentum, an essential part of them should be transmitted
from the lower part of the diagram,  Fig.3, to the upper one.
Nuclei of $^{238}$U have zero magnetic moment and can not capture
the monopoles.

\section{A new interaction}

In paper \cite{Baur}, changes of $\beta$-decay rate, with periods
24 hours and 27 days, were observed at two laboratories 140 km
apart. Extremum deviations of count rate ($ 0.7\%$ for $^{60}$Co
and $0.2\%$ for $^{137}$Cs) from the statistical average took
place for the both laboratories when they were oriented properly
along the three definite directions established in the outer
space. Bursts of count rate of beta-radioactive sources during
long-term measurements, similar to data of Ref.  \cite{Baur}, were
also reported in an independent paper \cite{Park}.

Series of papers devoted to a demonstration of a dependence of
$\alpha$-activity on cosmological factors was published by Prof.
S.E.~Shnol et al. (see, e.g., \cite{Shnoll1}, \cite{Shnoll2}). In
these studies, a phenomenon of a deviation of probability
distributions from the expected Poisson one was established. The
measurements were carried out in fixed with respect to the Earth's
surface laboratories during 5 minute time intervals.
Non-randomness of repetitions  of the shape of the observed
distributions was also established at the regular time intervals.
In short, the main results were the following:
\begin{enumerate}
\item Re-appearance of the same form of a probability distribution
took place most likely in the nearest  interval of observation.
\item There was a reliable growth of probability of the same form to re-appear
after 24 hours, 27 days, and one year.
\item Synchronous measurements of the form carried out in
different laboratories showed that for distances less than 100 km
about 60$\%$ pairs of the distributions had the same form.
Probability to observe similar distributions turned out to be high
also for measurements on a research ship in the Indian Ocean and
in a remote laboratory near Moscow, which were in the same time
zone.
\end{enumerate}

These data, in the case of their conformation, will almost
undoubtedly testify against the invariance of the radioactive atom
(and/or detector) properties with respect to spatial rotations.
According to the experiments of Shnoll at al., it is natural to
connect the observed effect with the influence of the nearest
cosmic environment, such as the Sun and the Moon. Authors of
\cite{Baur} explain a dependence of $\beta$-decay rate by the
mutual orientation of atoms and unknown cosmic field directed
toward the Constellation Hercules.

It is possible to give an explanation of the observed phenomena,
based on an idea that the decay rate depends on the atom
orientation with regards to some preferential direction in the
space. A concrete realization of this suggestion may be the
following.  Generators of the spinor representation of the
rotation sub-group,
\[ \sigma ^z  = \left( {\begin{array}{*{20}c}
   1 & 0  \\
   0 & { - 1}  \\
\end{array}} \right), \qquad \sigma ^ +  =  \sigma ^ x + i  \sigma ^ y   =
 \left( {\begin{array}{*{20}c}
   0 & 1  \\
   0 & 0  \\
\end{array}} \right), \qquad  \sigma ^ -  =  \sigma ^ x - i  \sigma ^ y  =
\left( {\begin{array}{*{20}c}
   0 & 0  \\
   1 & 0  \\
\end{array}} \right),
\]
can be factorized by means of the relations:
\[
\sigma ^ +   = a^\dag  b, \qquad  \sigma ^ -   = b^\dag  a,
\]
where
\[
a=|0\rangle \langle +|, \qquad b=|0\rangle \langle -|,
\]
and $|0\rangle $  is the vacuum state. Sign $^\dag$ denotes the
hermitian conjugation. Actually, we introduce in such a way the
birth and annihilation operators for atoms  capable  and incapable
of decay (atoms of the type $a$ and $b$, correspondingly). They
satisfy the fermion anticommutative relations,
\[
aa^\dag + a^\dag a = 1,  \qquad bb^\dag + b^\dag b = 1 ,
\]
which can be also interpreted as resolutions of identity in the
Fock spaces for particles of types  $a$ and $b$.

From the physical point of view, the undertaken factorization
implies the definition of  new quantum numbers, $$n_a =  a^\dag a,
\qquad n_b =  b^\dag b,$$ which correspond to probabilities of
atom to decay and to survive, correspondingly. Since $ \sigma ^z =
n_a - n_b$ and eigenvectors of operators $n_a$ and $n_b$ coincide
with eigenvectors of $ \sigma ^z $, it is evident that the atom
orientation which controls the radioactive atom decay is described
here in the close analogy with the description of spin 1/2
particles. Two different values of such a quasi-spin correspond to
atoms capable and incapable to decay.

Under spatial rotations, the ability of the system of the atom
plus the measuring instrument to demonstrate the decay, in the
general case, are changed: $$ Tr[\rho n_a] \rightarrow Tr[U \rho
U^\dagger \; V n_a V^\dagger].$$ Here $U_g$ and $V_g$ are unitary
operators describing transformations of an atom state, $\rho$, and
the measuring device under a spatial rotations $g$. The
probability distribution  will be invariant under the $g$
transformation iff $$ U_g = V_g .$$ In our consideration we
suggest that the measuring device does not change its properties
at spatial rotation, $ V_g = 1$.

If, e.g., we take an atom completely ready to decay, $ \rho_+ =
|+\rangle \langle +|, $ and rotate it relatively to the fixed
instrument, than corresponding transformations appear as follows:
$$ U\;|+\rangle = \alpha |+\rangle + \beta |-\rangle, $$ where
$|\alpha|^2 + |\beta|^2 =1.$ Thus the spatial rotations lead, in
our model, to changing the probability to observe the decay by the
factor $|\alpha|^2.$ We suggest that there are fixed directions in
the cosmic space such that atoms are the most unstable if their
quasi-spins are oriented along them.

A given source of radioactivity will demonstrate a dependence of
the decay rate on the orientation in the space only  if its
quasi-spin polarization is not equal to zero. However, the very
concept of the polarization implies that there is some interaction
which should orient atoms according to the minimum of their
energy. Thus, we come to a conclusion that the energy of an
unstable atom should depend on its orientation. This can be
described by inclusion into the Hamiltonian of the system a term
$$ H_0^{atom} = \frac{E}{2} \; \sum\limits_{i=1}^{N} \sigma_i^z
,$$ where summation is carried out over all radioactive atoms, and
$E$ is the energy differences between states with opposite
polarizations. The capability for decay of atoms can be changed by
an external field, $\varphi$, interacting with their
quasi-spins\footnote{It is clear that the energy of the
able-to-decay atom should be higher than the energy of one which
is unable. Therefore, quanta of $\varphi$ transmit some kind of
excitations. }. Our non-relativistic consideration does not forbid
us to introduce the following interaction in the spirit of the Lee
model \cite{Lee}: $$ H_{int} = \frac{\lambda}{\sqrt{N}}
\sum_{i=1}^N (\varphi a^\dagger_i b_i + b^\dagger_i a_i
\varphi^\dagger), $$ where $\lambda$ is a coupling constant. The
Hamiltonian of the field $\varphi$ has a usual form $$
H_0^{\varphi} = \hbar \omega \varphi^+ \varphi, $$ where $\omega$
is the frequency of quanta of the external field. The total
Hamiltonian, $$ H= H_0^{atom} + H_0^{\varphi} + H_{int}, $$
conserves the total number of atoms, $$ \sum\limits_{i=1}^{N}
a_i^+ a_i + b_i^+ b, $$ and the total ``readiness to decay'' $$
\frac{1}{2} \sum\limits_{i=1}^{N} (a_i^+ a_i - b_i^+ b_i) +
\varphi^+ \varphi .$$ In other words, quanta of $\varphi$  take
off and restore the capability of atoms to decay.

According to this model, experimentally observed variations of
nuclear decay rates could be a consequence of exchange between
radioactive atoms on the Earth and the Sun by quanta of the
$\varphi$ field (the corresponding Feynman graph is quite
obvious).

Interactions between radioactive atoms can be  also written in the
form of the Fermi 4-particle interaction, i.e. as
``current$\times$current'' \footnote{Here we use an analogy with
the four-fermion interaction which is a low energy approximation
for the second order diagram with the $W$ exchange. Quanta of
field $\varphi$ play here the role of $W$.}. As far as the current
components here are the Pauli matrices, $\sigma_x, \; \sigma_y,
\;\sigma_z$, an interaction invariant with regard to the spatial
rotations can be written in the form: $$ v = v(r) \; (\vec
\sigma_1 \cdot \vec \sigma_2). $$ It does not modify the total
quasi-spin,  $\vec S =\vec \sigma_1 + \vec \sigma_2 $, of two
interacting atoms. The obtained potential resembles the spin
dependent nucleon potential, and the theory of nuclear forces
prompts one more possible potential,
 $$  u = u(r) \;
[3(\vec \sigma_1 \cdot \vec n) (\vec \sigma_2 \cdot \vec n) -
(\vec \sigma_1 \cdot \vec \sigma_2) ],$$ which preserves  $\vec
S^2 .$  Apparently, assumptions of this model could be tested and,
if needed,  $v(r),$ $u(r)$, could be established  in space ship
experiments.

A current acting upon the radioactive atom may be not only the
quasi-spin of other atoms, but a pseudovector of a different
nature. In this connection, an assumption  that the pseudovector
of current of the  light monopole suggested by of G. Lochak can be
an effective catalyst of the weak decays is of interest.

We are grateful to G. Lochak for numerous interesting discourses
and to D.V. Filippov for a useful discussion of an influence of
the external magnetic field on probability of the $\beta$-decay.


\begin{thebibliography}{50}
%
\bibitem{Dirac} P.A.M.~Dirac. Proc. Roy. Soc. (London) A133 (1931)
60.
%
\bibitem{Lochak1} G.~Lochak. Ann. Fond. L. de Broglie 8 (1983)
345; 9 (1984) 5.
%
\bibitem{Lochak2} G.~Lochak.  Int. J. Theor. Phys.  24
(1985) 1019.
%
\bibitem{Lochak3} G.~Lochak. Z. Naturforsch. 62a (2007) 231.
%
\bibitem{Ryder} L. Ryder. Elementary Particles and Symmetries.
London: Gordon and Breach Science Publ., 1975.
%
\bibitem{Bjorken} J.D. Bjorken, S.D. Drell. Relativistic Quantum
Mechanics. New York: McGraw-Hill, 1964, 1965.
%
\bibitem{GSI1} M. Jung et al. Phys. Rev. Lett. 69 (1992) 2164.
%
\bibitem{GSI2}F. Bosch et al. Phys. Rev. Lett. 77 (1996) 5190.
%
\bibitem{Loudon} R. Loudon. Am. J. Phys. 27 (1959) 649.
%
\bibitem{one-dim} H. N. N\'u\~nez Y\'epez, C.A. Vargas, A.L. Salas
Brito. Eur. J. Phys. 8 (1987) 189.
%
\bibitem{Ruder} H. Ruder, H. Herold, W. R\"osner, G. Wunner.
Physica 127B (1984) 11-25.
%
\bibitem{Filippov} D.V. Filippov. Phys. Atom. Nucl. 70(2007)258.
%
\bibitem{Trubnikov} B.A.~Trubnikov. Theory of Plasma. Moscow:
Energoatomizdat, 1996 (in Russian).
%
\bibitem{Poincare}  H.~Poincar\'e. Comptes rendus Acad. Sc., 123
(1896) 530.
%
\bibitem{Mukhin} K.N. Mukhin Experimental Nuclear Physics, Vol. 1,
Moscow, Atomizdat, 1974 (in Russian).
%
\bibitem{Urutskoev} L.I. Urutskoev, D.V. Filippov. Physics -
Uspekhi 47 (2004) 1257.
%
\bibitem{Chernob} D.V. Filippov, L.I. Urutskoev, G. Lochak, A.A.
Rukhadze, in "Condensed Matter Nuclear Science", ed. Jean-Paul
Biberian. World Sientific, New Jersey-London-Singapore, 2006,
pp.838 -- 853.
%
\bibitem{Baur} Yu. A. Baurov, A.A. Konradov, V.F. Kushniruk, E.A.
Kuznetsov, Yu.G. Sobolev, Yu. V. Ryabov, A.P. Senkevich,  and S.V.
Zadorozsny. Mod. Phys. Lett. A 16 (2001) 2089.

\bibitem{Park}  A.G. Parkhomov,  Int. J. of Pure and Appl.
Phys.   1 (2005) 119.
%
\bibitem{Shnoll1}   S.E. Shnoll,  V.A. Kolombet,  E.V.  Pozharski,
T.A. Zenchenko,  I.M. Zvereva, and  A.A. Konradov. Physics-Uspehi
162 (1998) 1129.
%
\bibitem{Shnoll2}  S.E. Shnoll,  T.A. Zenchenko,  K.I. Zenchenko,
E.V. Pozharski,  V.A. Kolombet, and A.A. Konradov. Physics-Uspehi
43  (2000) 205 .
%
\bibitem{Lee} T.D.~Lee  Phys. Rev.  95 (1954) 1329.
%

\end{thebibliography}
\end{document}